\documentstyle[epsfig,epsf,amssymb]{EuroPhys}  

\newif\ifboo \boofalse

\newcommand{\nun}{\nu_{\rm HLL}}

\begin{document}  

\euro{}{}{}{} 

\Date{} 

\shorttitle{Chern-Simons Theory for Quantum Hall Stripes}

\title{Chern-Simons Theory for Quantum Hall Stripes}

\author{Bernd Rosenow$^{1,2}$ and Stefan Scheidl$^{2}$}

\institute{
  \inst{1} Physics Department, Harvard University,
  Cambridge, MA 02138, USA
  \\
  \inst{2} Institut f\"ur Theoretische Physik, Universit\"at zu
  K\"oln, D--50937 K\"oln, Germany} 

\rec{[draft \number\year/\number\month/\number\day/\number\time]}{} 

\pacs{\Pacs{73}{40.Hm}{Quantum Hall effect (integer and fractional)}
  \Pacs{71}{45.Lr}{Charge-density-wave systems}
  \Pacs{68}{65.+g}{Low-dimensional structures}}

\maketitle
  
\begin{abstract} 
  We develop a Chern-Simons theory to describe a two-dimensional
  electron gas in intermediate magnetic fields.  Within this approach,
  inhomogeneous states emerge in analogy to the intermediate state of
  a superconductor.  At half filling of the highest Landau level we
  find unidirectional charge-density-wave (CDW) solutions.  With a
  semiclassical calculation we give an intuitive explanation of the
  change of CDW orientation in the presence of an in-plane magnetic
  field.  An anisotropy in the electron band mass is suggested as a
  possible source of the reproducible orientation of the CDW.
\end{abstract} 


In two-dimensional electron systems both interactions and disorder
have surprising and striking consequences. A strong perpendicular
magnetic field quenches the kinetic energy and allows for the
observation of both fractional quantum Hall effect and composite
fermions in the lowest Landau level (LLL), where correlations among
electrons are especially important.  In higher Landau levels on the
other hand, interest has mainly been devoted to reentrant
localization-delocalization transitions causing the integer quantum
Hall effect, with interactions thought of as modifying its
non-universal features.  The discovery of a pronounced resistivity
anisotropy close to half filling of higher LLs
\cite{Lilly+99:a,DuTsui99} with filling factor $\nu > 4$ has recently
attracted attention to interaction effects in this regime. The
transport anisotropy is consistent with the formation of a
unidirectional CDW state which was predicted theoretically on the
basis of Hartree-Fock calculations \cite{FPA79,HF1,HF2}.  Numerical
exact diagonalization studies also support the idea of a CDW formation
\cite{ReHa99}. The orientation of this CDW structure is related to the
anisotropy observed in resistivity \cite{MacFi99,vOp99}.

In this letter, we show that a CDW ground state arises naturally in
the framework of a Chern-Simons (CS) theory \cite{Zhang92}.  The CS
approach successfully describes the fractional quantum Hall effect and
composite fermions in the LLL.  In intermediate LLs it is
complementary to Hartree-Fock calculations which become exact in the
limit of high Landau levels \cite{HF2}.  Our bosonic CS theory has
solutions similar to the intermediate state of superconductors and
allows for an intuitive interpretation of the striped phase: in the
highest, partially filled LL (HLL) domain walls separate
incompressible ``superconducting'' stripes with local filling factor
$\nun=1$ from empty stripes ($\nun=0$) penetrated by flux. The CDW
wave vector is determined by the competition of the domain-wall energy
favoring a long wave length and the Hartree energy which generally
increases with CDW wave length but has minima for special wave vectors
due to the modulation of the electron charge density by the
wave-function form factor.  The CDW wave length and structure are
calculated numerically within a mean-field approximation.
Furthermore, we calculate the wave-function form factor in the
presence of an in-plane magnetic field using a semiclassical
approximation. We find that for a thin quantum well the in-plane
component of the field is parallel to the axis of high resistivity, in
agreement with experiments \cite{Pan+99,Lilly+99:b} and numerical
calculations \cite{Jungwirth+99,StaMaPhi00}.  The band mass anisotropy
is shown to provide a mechanism determining the the {\it
  experimentally reproducible orientation} of the CDW wave vector for
perpendicular fields.


{\em Chern-Simons theory.} --- To employ the CS approach we first map
the problem of a partially filled $N$th LL onto an effective LLL
problem.  The scattering of electrons between different LLs can be
neglected when the Bohr radius $a_{\rm B}= (4 \pi \hbar^2 \epsilon_0
\epsilon_{\rm r})/(m e^2)$ is much smaller than the magnetic length
$\ell=\sqrt{\hbar/e B}$, a condition satisfied in experimentally
investigated systems.  We describe screening by the completely filled
LLs by a wave-vector dependent dielectric constant
$\epsilon(q)=\epsilon_0 \epsilon_{\rm r} \{1+(2/q a_{\rm B})[1-{\rm
  J}_0^2( q R_{\rm c})]\}$ \cite{Aleiner+95} with ${\rm J}_0$ denoting
the zeroth order Bessel function and $R_{\rm c}=\sqrt{2 N +1}\, \ell$
the cyclotron radius.  The intra--LL matrix elements of the Coulomb
interaction depend on the LL index $N$ only via the Laguerre
polynomial ${\rm L}_N$ in the form factor of the single particle
eigenstates \cite{interaction}. One obtains the same matrix elements
for states in the zeroth instead of the $N$-th LL but with the
effective interaction $V^{\rm eff}({\bf q})= {e^2 \over 2 
  \epsilon(q) q} {\rm L}_N^2(\ell^2 q^2 /2)$.  The Laguerre polynomial
can be interpreted as the quotient of the form factor of the $N$-th
and the zeroth LL as ${\rm L}_N=F_N/F_0$.  Since all LLs span
equivalent Hilbert spaces, the new interaction $V^{\rm eff}$ in the
LLL gives a faithful representation of the original problem.

The Fermi-Dirac statistics of electrons is mimicked by bosons carrying
one flux quantum $\Phi_0=h/e$ with them. The flux quanta are attached
by a singular gauge transformation \cite{Zhang92} with a statistical
gauge field ${\bf a}({\bf r})$ tied to the density of bosons $
\rho({\bf r})=\phi^\dagger({\bf r})\phi({\bf r})$ via the condition
$\mbox{\boldmath $\nabla$} \times {\bf a}({\bf r})= \Phi_0 \rho ({\bf
  r})$. The bosonic CS Hamiltonian is given by
\begin{equation} 
  {\cal H}= \int d^2 r \ \phi^\dagger({\bf r}) \frac 1{2m} 
  \left({-i \hbar} \mbox{\boldmath $\nabla$} + e \delta {\bf a}
  \right)^2 \phi({\bf r}) + 
  \frac 12 \int d^2r d^2r' \ \delta \rho({\bf r}) 
  V^{\rm eff}({\bf r}-{\bf r}') \delta \rho({\bf r}') \ ,
  \label{Hamiltonian}
\end{equation} 
with $\delta {\bf a}={\bf A +a}$, $\delta \rho =\Phi^\dagger \Phi -
\overline \nun B/ \Phi_0$, and $e$ the charge of the proton.  

{\em Mean-field analysis. }--- We perform a mean-field (MF) analysis
of the bosonic CS problem and replace the operators by a complex
field.  The kinetic energy (of the bosons, not to be confused with the
quenched kinetic energy of the electrons) alone would be minimized by
the homogeneous saddle points $\phi\equiv \sqrt{B / \Phi_0}$, $\delta
{\bf a}\equiv 0$ and $\phi \equiv 0$, $\delta {\bf a}\equiv {\bf A}$
corresponding to a completely filled or empty LL, respectively.
Without the interaction potential the ground state of a half filled LL
would clearly be one domain with filling factor $\nun = 1$ and another
domain with $\nun =0$.  However, due to the long-range nature of the
Coulomb interaction, these large domains break up into smaller ones in
such a way that the sum of domain-wall and Coulomb energy is minimal,
and a charge-density wave is formed in analogy to the intermediate
state of a superconductor.  The effective field $\mbox{\boldmath
  $\nabla$} \times \delta {\bf a}$ is expelled from the
``superconducting'' regions with $\nun = 1$.  For average fillings
$\overline \nun=1 -\epsilon$ close to one on the other hand, the
solution with lowest energy is given by a flux-tube array in analogy
to phases of type-I superconductors in the intermediate state, where
each flux tube can carry a single or multiple flux quanta.  For the
electron problem this is equivalent to the formation of a Wigner
crystal.

To make the physics of the mean-field solution transparent, we express
the wave function by amplitude and phase, introduce a gauge invariant
``velocity'' ${\bf Q}$, and express the boson current density ${\bf
  j}$ in these new variables
\begin{eqnarray} 
  \phi({\bf r}) = \sqrt{\frac{B}{\Phi_0}} f({\bf r}) e^{i \theta({\bf 
      r})}, \quad 
  {\bf Q}({\bf r})=\ell^2 \left( \mbox{\boldmath$ \nabla$} \theta  + 
    \frac{2 \pi}{\Phi_0} \delta{\bf a} \right), \quad
  {\bf j}({\bf r})={- e \omega_{\rm c}\over 2 \pi\ell^2}\, 
  f^2\, {\bf Q}({\bf r})\ .  
\end{eqnarray} 
 
\begin{figure}
  \centering
  \epsfig{file=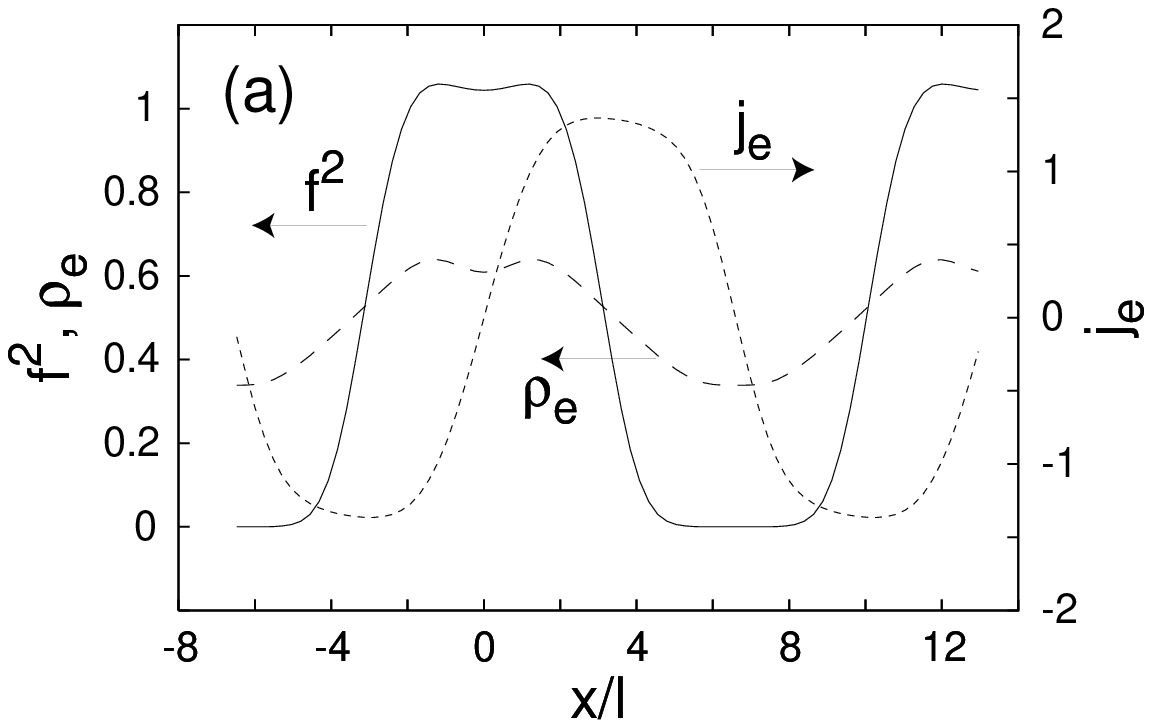,width=7.1cm}
  \epsfig{file=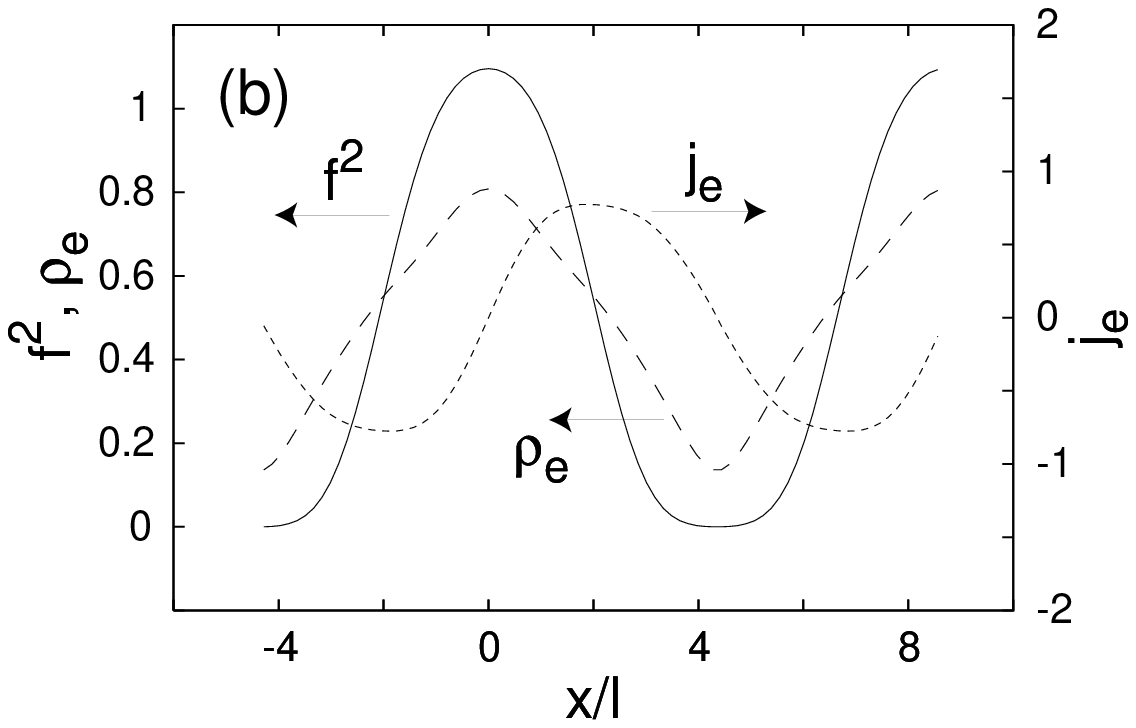,width=7.1cm}
  \caption{Structure of the MF solution: 
    local filling factor of bosons $f^2$ (full line), local filling
    factor of electrons $\rho_e$ (dashed line; the charge density is $-e
    \rho_e B/\Phi_0$), and charge-current density $j_e$ (dotted line; in
    arbitrary units) as function of $x/\ell$ (a) for $N=8$ with the
    optimal period $\Lambda \approx 3.1 R_{\rm c} \approx 12.8\ell$ and
    (b) for $N=2$ with the optimal period $\Lambda \approx 3.9 R_{\rm c}
    \approx 8.7\ell$.}
  \label{fig.struct}
\end{figure}

We enforce the constraint on ${\bf a}({\bf r})$ with the help of a
Lagrange multiplier $\lambda$ (which is the rescaled zero component of
the CS field) and find the saddle-point equations
\begin{eqnarray} 
  -f^{\prime\prime} + {1\over\ell^4} {\bf Q}^2 f 
  + 2 (u_{\rm H}-\lambda) f=0\ ,\ \ \ \  
  \mbox{\boldmath $\nabla$} \times {\bf Q} 
  = (v+ f^2 - 1)\hat{\bf z}\ \ , \ \ \ 
  \mbox{\boldmath $\nabla$} \lambda 
  =  {1\over\ell^4} f^2  {\bf Q} \times 
  \hat{\bf z}
  \label{saddlepoint}
\end{eqnarray} 
with the vortex density ${\bf v}({\bf r})=\ell^2 \mbox{\boldmath
  $\nabla$} \times \mbox{\boldmath $\nabla$}\theta({\bf r})$. The
Hartree potential $u_{\rm H}$ is expressed in units of $\ell^2 \hbar
\omega_{\rm c}$.  One immediate consequence of the saddle-point
equations is the screening of an electrostatic potential by the CS
superconductor in analogy to the screening of a vector potential by a
conventional superconductor. A negative potential causes a reduction
of $\nun=f^2$ from one and induces a current via the second part of
Eq.~(\ref{saddlepoint}).  This argument can be made quantitative by
describing the current response to an electric field with the help of
the Hall conductivity $\sigma_{\rm H}= e^2/h$ as ${\bf j}= \sigma_{\rm
  H} {\bf E}\times \hat{\bf z}$.  From the third part of
Eq.~(\ref{saddlepoint}) one finds $\mbox{\boldmath $\nabla$} \lambda =
(e / \hbar \omega_{\rm c} \ell^2) {\bf E}$ leading to an exact
cancellation of the external potential in the Schr\"odinger equation
for the superfluid density $f^2$. Via this mechanism a Bose
condensation in a spatially varying external potential is possible.
Note that the bosonic wave functions are not restricted to the LLL in
the variational approach.  This is an valid approximation in the limit
where the electron-electron interaction is weak compared to the
inter-Landau-level spacing \cite{Aleiner+95} and a mixing of Landau
levels is negligible.  The same assumption is used also in the
Hartree-Fock approaches \cite{HF1,HF2} in order to justify that the
ground state does not involve single-particle states above the
partially filled highest LL.  Since on the scale $\ell$ of the typical
particle distance the strength of the bosonic interaction $V^{\rm
  eff}$ is comparable to that of the original fermionic interaction
(for $N$ not too large as in experiments), the ground state of the
bosonic problem (\ref{Hamiltonian}) will lie to a good approximation
in the LLL, even if we search in an extended space of functions
including higher LLs.  Dropping the variational constraint in the
bosonic problem therefore induces inaccuracies of the same order of
magnitude as ignoring the Landau-level mixing in the fermionic
problem. As can be seen from the variational result below (Fig.
\ref{fig.struct}), the local filling factor overshoots only by a few
percent over one, which indicates that the solution lies to a very
good approximation within the LLL although the constraint is not
imposed.


{\em 1D numerical solution.} --- In order to confirm the physical
picture developed above and to provide a comparison of our CS approach
with previous Hartree-Fock approaches \cite{HF1,HF2}, we have
numerically determined the MF solution at half filling ($\overline
\nun =\frac 12$).  This was done approximately by assuming a
unidirectional modulation, i.e., that the charge density is modulated
with a period $\Lambda$ in $x$ direction and is constant in $y$
direction.  In this approximation, vortices cannot be taken into
account as point-like objects.  Instead, all vortices within a period
have to be concentrated on lines, $v({\bf r})= \sum_n \frac \Lambda 2
\delta(x-(n+1/2)\Lambda)$.  For symmetry reasons, currents then flow
only in $y$ direction, ${\bf Q}({\bf r})=Q_y(x) {\bf e}_y$. To obtain
the MF solution, we use a variational ansatz $f(x)=\sum_k f_k \cos(2
\pi k x/\Lambda)$ with the constraint $f((n+ \frac 12) \Lambda)=0$ at
the vortex positions.  In a first step, we minimize ${\cal H}$ for
given $\Lambda$ by varying the coefficients $f_k$ with the constraints
of half filling and $\partial_x Q_y(x)=v+f^2-1$. Then the resulting
energy ${\cal H}(\Lambda)$ is minimized as a function of $\Lambda$ to
obtain the CDW period. The choice of $a_{\rm B}=\ell/\sqrt \nu$ fixes
the interaction strength and is representative for experiments
\cite{Lilly+99:a,Pan+99}. A full 2D numerical solution and the analysis
of gauge field fluctuations \cite{{Zhang92}} is beyond the scope of the 
present analysis.  

We present our results for filling factors $\nu=2N+\frac 12$ in the
Landau levels $N=8$ and $N=2$.  The structure of the MF solutions is
depicted in Fig.  \ref{fig.struct}. For $N=8$, $f^2$ shows a clear
separation between full ($f^2 \approx 1$) and empty ($f^2\approx 0$)
regions.  The width of the transition between the full and empty
region is of order $\ell$.  Therefore, the separation between full and
empty regions is blurred out for smaller $N$ (for $N=2$, see Fig.
\ref{fig.struct}b).  The actual charge and charge-current density,
which are obtained after convoluting $f^2$ and $f^2 Q_y$ with the
appropriate form factors are much smoother than $f^2$ and vary only on
the scale of $R_{\rm c}$.  The charge density within this LL shows a
relative variation of about 20\% for $N=8$.  This magnitude is roughly
consistent with Hartree-Fock calculations \cite{HF1,HF2} for large
$N$. For $N=2$ the charge density shows a stronger modulation of about
60\%.  Thus, although charge and charge-current density are modulated
only on the scale $\Lambda$ for all $N$, the relative variation of
charge density is found to increase with decreasing $N$.

\begin{figure}
  \centering
  \epsfig{file=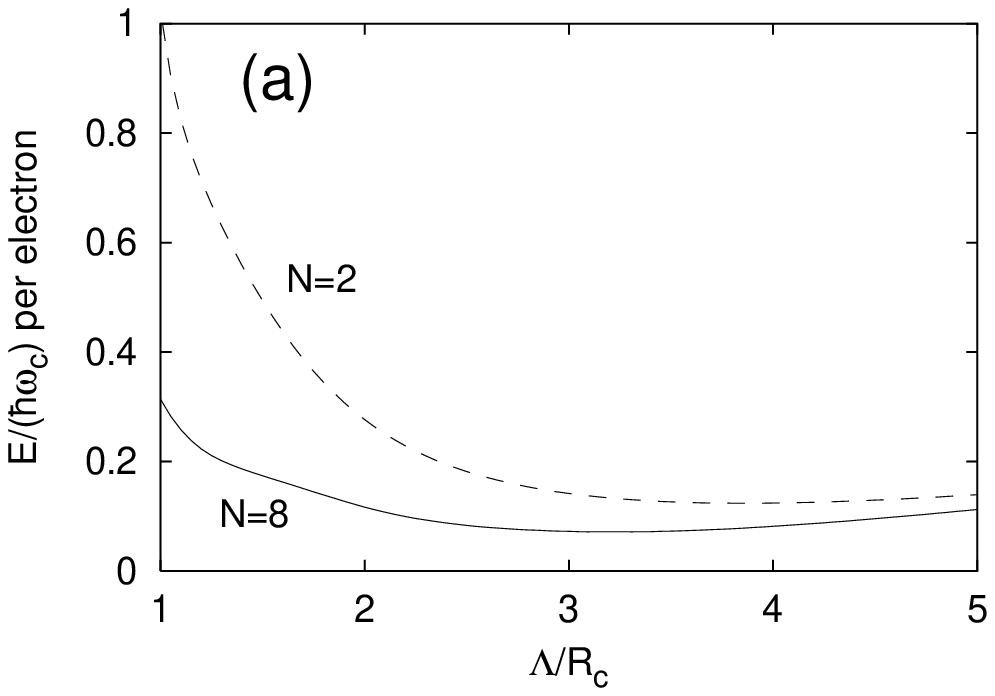,height=5cm}
  \epsfig{file=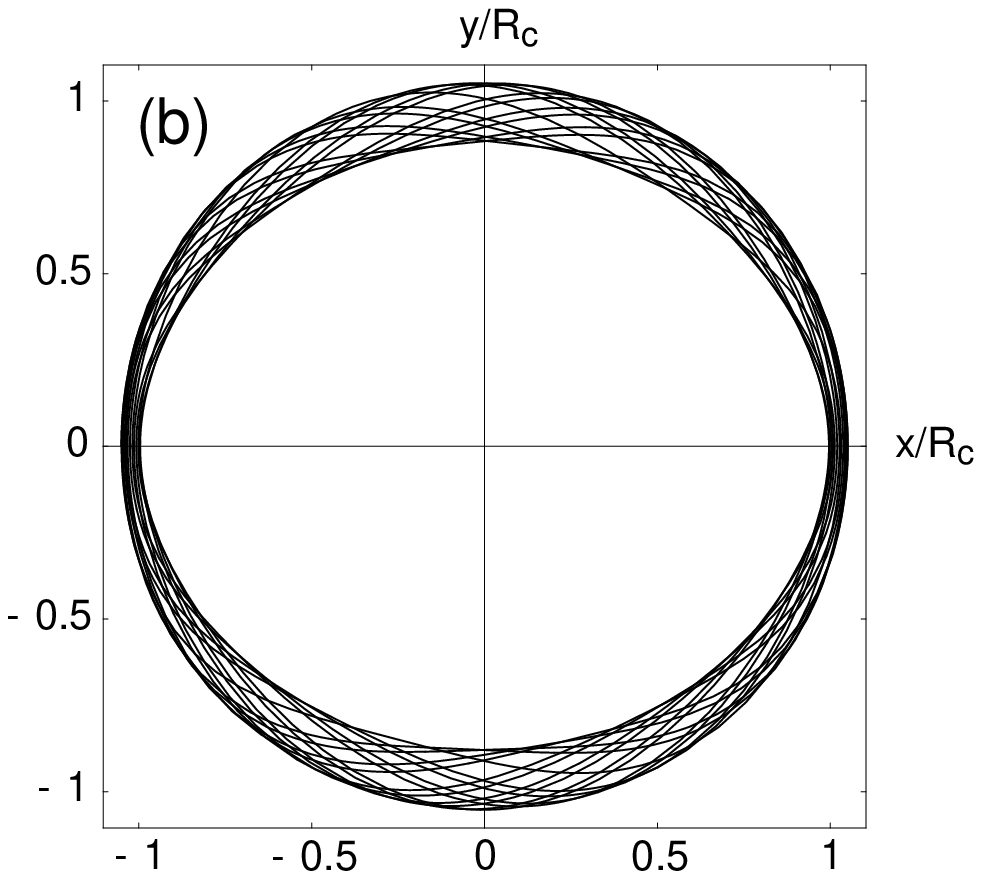,width=6cm}
  \caption{(a) Plot of the energy per electron in units of 
    $\hbar \omega_{\rm c}$ as a function of $\Lambda/R_{\rm c}$ for
    $N=8$ (full line) and $N=2$ (dashed line). (b) Plot of an orbit for
    a field tilted by 45$^\circ$ with respect to the $xy$ plane (with
    $N=2$, $\omega_x=\omega_z$ and $\omega_0=3 \omega_z$ over a time of
    about 16$T_-$). Lengths are in units of $R_{\rm c}$. The cyclotron
    mode has its long axis parallel to the in-plane field (in $x$
    direction), whereas the oscillator mode has its longer axis in $y$
    direction. }
  \label{fig.E_L}
\end{figure}

The energy per electron as a function of the periodicity is displayed
in Fig.  \ref{fig.E_L}.  The energy scale in our saddle-point
approximation is the cyclotron energy.  We expect that gauge-field
fluctuations will renormalize it to the Coulomb energy $e^2/4 \pi
\epsilon_0 \epsilon_{\rm r} \ell$. The optimal periodicity
$\Lambda_N$, given by the global minimum of this function, is found at
$\Lambda_8 \approx 3.1 R_{\rm c}$ and $\Lambda_{2} \approx 3.9 R_{\rm
  c}$.  Thus, the periodicity is larger than expected from the first
zero of the form factor at $\Lambda \approx 2.6 R_{\rm c}$ in
agreement with HF calculations \cite{HF1,HF2}. However, we find that
$\Lambda_N/R_{\rm c}$ increases notably with decreasing $N$.  In our
approach this effect can be understood from the increasing weight of
the ``kinetic'' or domain-wall contribution within the CS energy.

{\em Tilted magnetic field.} --- So far, we have studied the formation
of a CDW by electrons in high Landau levels within the CS approach.
In experiments \cite{Lilly+99:a,Lilly+99:b} the longitudinal
resistivity appears to be systematically higher in $[1\bar10]$
directions as compared to $[110]$ directions.  This raises the
question of what mechanism breaks the fourfold crystal symmetry and
orienting the CDW wave vector parallel to the $[1\bar10]$ directions.
For this reason resistivity was studied in {\em tilted} magnetic
fields \cite{Pan+99,Lilly+99:b} and a sufficiently strong in-plane
component ${\bf B}_\parallel$ of the magnetic field was found to
orient the directions of large resistivity parallel to ${\bf
  B}_\parallel$.  We present a mechanism for the coupling between the
CDW wave vector and ${\bf B}_\parallel$ that involves the dynamics of
the electrons in the $z$ direction perpendicular to the plane. We
model the confinement of the electrons in this direction by a harmonic
potential with eigenfrequency $\omega_0$.  While similar models were
studied recently in the framework of a Hartree-Fock approach
\cite{Jungwirth+99,StaMaPhi00} and of Laughlin-like states
\cite{Yu+99}, we use a simple semiclassical approach to obtain the
anisotropic form factor of cyclotron orbits in tilted fields, which
then can easily be combined with the theory of untilted fields.
 
The classical equations of motion for a single electron subject to a
tilted magnetic field ${\bf B}=(B_\parallel,0,B_z)$ and a harmonic
confining potential are linear and possess two eigenmodes with
frequencies $ \omega_\pm^2 = \frac 12
(\omega_0^2+\omega_z^2+\omega_x^2) \pm \frac 12
\sqrt{(\omega_0^2+\omega_z^2+\omega_x^2)^2 - 4 \omega_0^2
  \omega_z^2}$, where $\omega_x$ and $\omega_z$ are the cyclotron
frequencies associated with the in-plane and out-of-plane components
of the magnetic field, which typically are small compared to
$\omega_0$.  For vanishing $B_\parallel$, the modes with frequency
$\omega_+^2 \approx \omega_0^2 [1 + \omega_x^2 /
(\omega_0^2-\omega_z^2)]$ or $\omega_-^2 \approx \omega_z^2 [1 -
\omega_x^2 / (\omega_0^2-\omega_z^2)]$ become a pure oscillation in
$z$ direction or a cyclotron orbit in the $xy$ plane, respectively.
We determine the semiclassical orbits from the Bohr quantization
condition $\oint {\bf p}_\sigma \cdot {\bf r}_\sigma=2 \pi \hbar
(\frac 12 + n_\sigma)$, where $\sigma=\pm$ and we choose the ground
state, $n_+=0$, for the oscillator-like mode and the and the $N$th
Landau level, $n_-=N$, for the cyclotron-like mode. This choice is
unique for $\omega_0 > \omega_z$.  Subsequently, we consider only the
projection of the orbits onto the $xy$ plane, which are ellipses with
half axes of lengths
\begin{eqnarray} 
  R_{x \sigma }^2 =(1+2N_\sigma) \frac \hbar{m \omega_\sigma} 
  \frac{\omega_z^2} {\omega_z^2 + \omega_x^2 
    \omega_\sigma^4/(\omega_0^2-\omega_\sigma^2)^2}\ , \quad R_{y \sigma 
   }^2 = \frac {\omega_\sigma^2}{\omega_z^2} R_{x \sigma}^2\ . 
\end{eqnarray} 
The cyclotron-like ellipse is larger than the oscillator-like ellipse
($R_{\alpha -} > R_{\alpha +}$). In addition, the cyclotron-like
ellipse is longer in $x$ direction than in $y$ direction
($R_{x-}>R_{y-}$), whereas for the oscillator-like ellipse the
opposite holds true (cf. Fig.  \ref{fig.E_L}b).  Because of this
contrary deformation the resulting anisotropy of the form factor is
delicate.  The form factor $F({\bf q})$ is the Fourier transform of
the density $\rho({\bf r})$ of finding the electron at a position
${\bf r}={\bf r}_-(t)+{\bf r}_+(t)$.  It factorizes into two
contributions from the normal modes, $F({\bf q})=F_-({\bf q}) F_
+({\bf q})$, which are the Fourier transforms of the density
$\rho_\pm({\bf r})= \frac1 {T_\pm} \int_0^{T_\pm} dt \delta({\bf
  r}-{\bf r}_\pm(t))$ of the normal mode orbits, where $T_\pm = 2
\pi/\omega_\pm$ is the orbit period.
 
In the presence of the tilted field the form factor is anisotropic and
we now discuss the location of its first zero at small momenta, which
essentially determines the CDW periodicity and energy.  At small
momenta, the zeros of $F$ are those of $F_-$ since the cyclotron-like
orbits have a larger radius.  If one estimates the CDW periodicity
from the location of the zeros of $F_-({\bf q})$ one finds the
periodicity proportional to the radius of the cyclotron ellipse in the
direction of the CDW wave vector.  The total energy of the CDW is then
minimized if its wave vector points parallel to the long direction of
the cyclotron ellipse, i.e., parallel to ${\bf B}_\parallel$.  This
conclusion is in agreement with experiments \cite{Pan+99,Lilly+99:b}.

Finally, we estimate the energy difference between an orientation of
the CDW wave vector parallel and perpendicular to the in-plane field
in the LL $N=2$ (assuming again $\omega_x \lesssim \omega_z \ll
\omega_0$).  The energy change is related to a change of the CDW
period which can be approximated by $\delta \Lambda /\Lambda \approx
(R_{x-}/R_{\rm c} -1) \approx (\omega_x/\omega_0)^2$.  The change of
the energy per particle with $\Lambda$ is estimated numerically as
$\delta E \approx - 0.1 \ \hbar \omega_{\rm c} \ \delta \Lambda
/\Lambda $ from the domain-wall energy.  The tilt energy per particle
is then given by
$ 
 E_{\rm tilt}= \tan^2(\theta) \left({\omega_z / \omega_0}\right)^2
  3 {\rm K}
$
>From the threshold angle $\theta \approx 10^\circ$ \cite{Lilly+99:b}
where the CDW orientation follows the in-plane component of the field
and a typical ratio $\omega_z \sim 0.15\, \omega_0$ we find $E_{\rm
  tilt} \approx 1$mK.  Consequently, the intrinsic symmetry breaking
mechanism in the absence of a tilt must lead to an energy gain of the
same magnitude. We now examine two possible mechanisms.


{\em Anisotropy.} --- In the normal state the electron gas has
anisotropic resistivity with ``hard'' axis parallel to the $[110]$
crystallographic direction, i.e., perpendicular to the hard axis
observed in the CDW state.  This normal anisotropy, which has been
reported for some of the samples studied in \cite{Lilly+99:a}, can be
ascribed to scattering processes caused by an interface roughness with
anisotropic correlation lengths $\xi_{[1\bar10]} > \xi_{[110]}$
\cite{ToSaTaHo92}.  The observed surface morphology \cite{Willett+00}
is consistent with such an anisotropic interface roughness.  Since the
anisotropy in the CDW state is determined by collective pinning,
whereas it is determined by single-electron scattering in the normal
state, the direction of the hard axis is not necessarily identical and
therefore we now examine the collective pinning mechanism.  Variations
of the interface height $d$ result in an electrostatic potential $
\phi_d({\bf q})=\frac 12 e n d({\bf q})/ \epsilon(q) $.  Here $n$ is
the density of positive donors and the height fluctuations are assumed
to be Gaussian correlated, $[|d({\bf q})|^2]_{\rm av}= 2 \pi d_0^2
\xi_x \xi_y \exp[-(q_x^2 \xi_x^2 + q_y^2 \xi_y^2)/2]$ with correlation
lengths $\xi_x$, $\xi_y \gg \Lambda$ \cite{Willett+00} and a typical
height $d_0 \approx 2.8${\AA} of the order of a GaAs monolayer.
Pinning leads to an energy gain via local displacements of the CDW
profile which are governed by a smectic elastic response
\cite{MacFi99}.  From the curvature of the CDW energy as a function of
the wave length we estimate the compression modulus $K_x \approx
10^{11}$eVm$^{-2}$ and the bending modulus $K_y \approx
(\Lambda/4\pi)^2 K_x$.  For parameters representing the experiments
\cite{Lilly+99:a,DuTsui99} we estimated the pinning energy density in
a linear response calculation \cite{cutoff},
\begin{eqnarray}
  E_{\rm rough} \approx -  \left(\frac {n \pi e}{2 \nu \Lambda}
  \right)^2  
  \int {d^2 q \over (2 \pi)^2}  
  \frac{\left[ |\phi_d (q_x+2 \pi/\Lambda, q_y)|^2\right]_{\rm av}}
  {K_x q_x^2 + K_y q_y^4}.
\end{eqnarray}
Note that the pinning strength is given by the disorder correlator
near the wave vector $2\pi/\Lambda$ and therefore it is exponentially
small, $\left[ |\phi (q_x+2 \pi/\Lambda, q_y)|^2\right]_{\rm av}
\propto e^{- 2 (\pi \xi_x/\Lambda)^2}$.  For $\xi_x \gg \Lambda$ this
exponential dependence suppresses the pinning energy per particle by
orders of magnitude below 1mK per particle as determined above from
the threshold tilt angle.  Therefore interface roughness is irrelevant
for the CDW orientation. In addition it would imply that the CDW wave
vector is aligned in the direction of the shorter disorder correlation
length, i.e., the $[110]$ direction \cite{ToSaTaHo92,Willett+00} in
contradiction to experiments.

As pointed out in Ref.  \cite{Kro99} the electric field $E=e n/2
\epsilon_0 \epsilon_{\rm r}$ in $[00\bar1]$ direction between the
electron system and the donor layer provides a different symmetry
breaking mechanism.  We now show that this field generates an
anisotropic electronic band mass which induces the observed CDW
orientation by an energy gain on the appropriate scale.  Since Ga and
As carry opposite partial charges they are displaced such that the
GaAs bonds in $[11\cdot]$ direction are stretched whereas the bonds in
$[1\bar1\cdot]$ direction are shortened.  From pressure experiments
\cite{mass} it is known that a shortening of bonds leads to an
increase of the effective band mass $m^*$. We estimate the effective
mass changes $m_{[1\bar10]/[110]}^*/m^*-1 \approx \pm 10^{-4} $ by
identifying the electric field with an effective pressure
\cite{pressure}. This anisotropic band mass affects the CDW via the
form factor.  A semiclassical analysis implies elliptic cyclotron
orbits with half axes $R_{[1\bar10]/[110]}/R_c-1 \sim \pm 10^{-4}$
which leads to an increased CDW period $\delta
\Lambda_{[1\bar10]/[110]}/\Lambda \approx \pm 10^{-4}$.  In analogy to
the analysis of the in-plane field this implies that the mass
anisotropy prefers the $[1\bar10]$ orientation over the $[110]$
orientation by about 1mK per electron in agreement with the
experimentally observed orientation and the energy scale derived from
the tilt experiments.

{\em Conclusion.} --- We have developed a bosonic Chern-Simons theory
to investigate the formation of a CDW state in intermediate Landau
levels and the influence of an anisotropic interface potential on the
CDW orientation. In a semiclassical calculation we have found that the
CDW wave vector is parallel to an in-plane magnetic field for quantum
wells with a hard confining potential.  The band mass anisotropy is a
possible origin for the CDW orientation in perpendicular fields.

{\em Acknowledgments.} --- We are grateful to B. I. Halperin, M. M.
Fogler, M. P.  Lilly, R. H. Morf, and F. von Oppen for stimulating
discussions and thank A. Kleinschmidt for assistance during the
initial stages of the numerical part of this work. B.R. was supported
by DFG grant Ro2247/1-1.


\begin{thebibliography}{99}
\bibitem{interaction} The matrix element of the Coulomb interaction
  between eigenstates in Landau gauge is given by $\langle X_1 X_2
  |\hat{V}| X_3 X_4 \rangle = (e^2/4\pi \epsilon L^2_y)
  \delta(X_1-X_3-X_2+X_4) \int d^2 q \ \delta (q_y+(X_1-X_3)/\ell^2)
  \exp \left[(X_1+ X_3-iq_x \ell^2)^2/4 \ell^2 + (X_2+X_4-i q_x
    \ell^2)^2/4 \ell^2\right] {\rm L}^2_N \left(\ell^2 q^2/2\right) 1/
  q$. Note that it depends on the Landau level number $N$ only through
  the Laguerre polynomial.
\bibitem{cutoff} This integral is IR divergent and has to be cut off
  by the Larkin length where the linear response treatment of the
  pinning energy breaks down.  For weak disorder this length coincides
  wih the system size.
\bibitem{pressure} The donor layer generates a homogeneous electric
  field $E=e n/2 \epsilon_0 \epsilon_{\rm r}$, which exerts a pressure
  $p=E e^{*}/a_0^2\approx 2\times 10^5$Pa on the sublattice of Ga
  atoms relative to the sublattice of Al atoms. Here $n=2\cdot
  10^{15}$ is the density of donors, $e^*\approx 0.3 e$ the partial
  charge of ligands, and $a_0=0.56$nm the distance between likewise
  atoms.
\end{thebibliography}
\end{document}